\RequirePackage{lineno}  

\documentclass[aps,prl,twocolumn,superscriptaddress,showpacs]{revtex4} 

\usepackage{graphicx}
\usepackage{rotating}

\begin{document}

\thispagestyle{empty}


\vspace{0.5in}


\newcommand{\fix}[1]{} 
\newcommand{\MET}{\mbox{$\raisebox{.3ex}{$\not\!$}E_T$}}
\newcommand{\METVEC}{\mbox{$\raisebox{.3ex}{$\not\!$}{\vec E}_T$}}
\newcommand{\MPTVEC}{\mbox{$\raisebox{.3ex}{$\not\!$}{\vec {p}_T}$}}

\def\DZero{D0~}

\newcommand{\Et}{\mbox{$E_T$}}
\newcommand{\Pt}{\mbox{$p_T$}}
\newcommand{\Ht}{\mbox{$H_T$}}
\newcommand{\met}{\mbox{$\protect \raisebox{0.3ex}{$\not$}\Et$}}
\newcommand{\mpt}{\mbox{$\protect \raisebox{0.3ex}{$\not$}\Pt$}}



\title{Invariant Mass Distribution of Jet Pairs Produced in Association \\ with a \boldmath{$W$} boson in \boldmath{$p \bar{p}$} Collisions at \boldmath{$\sqrt{s}$} = 1.96 TeV}
\vspace{0.5in}

\affiliation{Institute of Physics, Academia Sinica, Taipei, Taiwan 11529, Republic of China} 
\affiliation{Argonne National Laboratory, Argonne, Illinois 60439, USA} 
\affiliation{University of Athens, 157 71 Athens, Greece} 
\affiliation{Institut de Fisica d'Altes Energies, ICREA, Universitat Autonoma de Barcelona, E-08193, Bellaterra (Barcelona), Spain} 
\affiliation{Baylor University, Waco, Texas 76798, USA} 
\affiliation{Istituto Nazionale di Fisica Nucleare Bologna, $^{aa}$University of Bologna, I-40127 Bologna, Italy} 
\affiliation{University of California, Davis, Davis, California 95616, USA} 
\affiliation{University of California, Los Angeles, Los Angeles, California 90024, USA} 
\affiliation{Instituto de Fisica de Cantabria, CSIC-University of Cantabria, 39005 Santander, Spain} 
\affiliation{Carnegie Mellon University, Pittsburgh, Pennsylvania 15213, USA} 
\affiliation{Enrico Fermi Institute, University of Chicago, Chicago, Illinois 60637, USA}
\affiliation{Comenius University, 842 48 Bratislava, Slovakia; Institute of Experimental Physics, 040 01 Kosice, Slovakia} 
\affiliation{Joint Institute for Nuclear Research, RU-141980 Dubna, Russia} 
\affiliation{Duke University, Durham, North Carolina 27708, USA} 
\affiliation{Fermi National Accelerator Laboratory, Batavia, Illinois 60510, USA} 
\affiliation{University of Florida, Gainesville, Florida 32611, USA} 
\affiliation{Laboratori Nazionali di Frascati, Istituto Nazionale di Fisica Nucleare, I-00044 Frascati, Italy} 
\affiliation{University of Geneva, CH-1211 Geneva 4, Switzerland} 
\affiliation{Glasgow University, Glasgow G12 8QQ, United Kingdom} 
\affiliation{Harvard University, Cambridge, Massachusetts 02138, USA} 
\affiliation{Division of High Energy Physics, Department of Physics, University of Helsinki and Helsinki Institute of Physics, FIN-00014, Helsinki, Finland} 
\affiliation{University of Illinois, Urbana, Illinois 61801, USA} 
\affiliation{The Johns Hopkins University, Baltimore, Maryland 21218, USA} 
\affiliation{Institut f\"{u}r Experimentelle Kernphysik, Karlsruhe Institute of Technology, D-76131 Karlsruhe, Germany} 
\affiliation{Center for High Energy Physics: Kyungpook National University, Daegu 702-701, Korea; Seoul National University, Seoul 151-742, Korea; Sungkyunkwan University, Suwon 440-746, Korea; Korea Institute of Science and Technology Information, Daejeon 305-806, Korea; Chonnam National University, Gwangju 500-757, Korea; Chonbuk National University, Jeonju 561-756, Korea} 
\affiliation{Ernest Orlando Lawrence Berkeley National Laboratory, Berkeley, California 94720, USA} 
\affiliation{University of Liverpool, Liverpool L69 7ZE, United Kingdom} 
\affiliation{University College London, London WC1E 6BT, United Kingdom} 
\affiliation{Centro de Investigaciones Energeticas Medioambientales y Tecnologicas, E-28040 Madrid, Spain} 
\affiliation{Massachusetts Institute of Technology, Cambridge, Massachusetts 02139, USA} 
\affiliation{Institute of Particle Physics: McGill University, Montr\'{e}al, Qu\'{e}bec, Canada H3A~2T8; Simon Fraser University, Burnaby, British Columbia, Canada V5A~1S6; University of Toronto, Toronto, Ontario, Canada M5S~1A7; and TRIUMF, Vancouver, British Columbia, Canada V6T~2A3} 
\affiliation{University of Michigan, Ann Arbor, Michigan 48109, USA} 
\affiliation{Michigan State University, East Lansing, Michigan 48824, USA}
\affiliation{Institution for Theoretical and Experimental Physics, ITEP, Moscow 117259, Russia}
\affiliation{University of New Mexico, Albuquerque, New Mexico 87131, USA} 
\affiliation{Northwestern University, Evanston, Illinois 60208, USA} 
\affiliation{The Ohio State University, Columbus, Ohio 43210, USA} 
\affiliation{Okayama University, Okayama 700-8530, Japan} 
\affiliation{Osaka City University, Osaka 588, Japan} 
\affiliation{University of Oxford, Oxford OX1 3RH, United Kingdom} 
\affiliation{Istituto Nazionale di Fisica Nucleare, Sezione di Padova-Trento, $^{bb}$University of Padova, I-35131 Padova, Italy} 
\affiliation{LPNHE, Universite Pierre et Marie Curie/IN2P3-CNRS, UMR7585, Paris, F-75252 France} 
\affiliation{University of Pennsylvania, Philadelphia, Pennsylvania 19104, USA}
\affiliation{Istituto Nazionale di Fisica Nucleare Pisa, $^{cc}$University of Pisa, $^{dd}$University of Siena and $^{ee}$Scuola Normale Superiore, I-56127 Pisa, Italy} 
\affiliation{University of Pittsburgh, Pittsburgh, Pennsylvania 15260, USA} 
\affiliation{Purdue University, West Lafayette, Indiana 47907, USA} 
\affiliation{University of Rochester, Rochester, New York 14627, USA} 
\affiliation{The Rockefeller University, New York, New York 10065, USA} 
\affiliation{Istituto Nazionale di Fisica Nucleare, Sezione di Roma 1, $^{ff}$Sapienza Universit\`{a} di Roma, I-00185 Roma, Italy} 

\affiliation{Rutgers University, Piscataway, New Jersey 08855, USA} 
\affiliation{Texas A\&M University, College Station, Texas 77843, USA} 
\affiliation{Istituto Nazionale di Fisica Nucleare Trieste/Udine, I-34100 Trieste, $^{gg}$University of Trieste/Udine, I-33100 Udine, Italy} 
\affiliation{University of Tsukuba, Tsukuba, Ibaraki 305, Japan} 
\affiliation{Tufts University, Medford, Massachusetts 02155, USA} 
\affiliation{University of Virginia, Charlottesville, VA  22906, USA}
\affiliation{Waseda University, Tokyo 169, Japan} 
\affiliation{Wayne State University, Detroit, Michigan 48201, USA} 
\affiliation{University of Wisconsin, Madison, Wisconsin 53706, USA} 
\affiliation{Yale University, New Haven, Connecticut 06520, USA} 
\author{T.~Aaltonen}
\affiliation{Division of High Energy Physics, Department of Physics, University of Helsinki and Helsinki Institute of Physics, FIN-00014, Helsinki, Finland}
\author{B.~\'{A}lvarez~Gonz\'{a}lez$^w$}
\affiliation{Instituto de Fisica de Cantabria, CSIC-University of Cantabria, 39005 Santander, Spain}
\author{S.~Amerio}
\affiliation{Istituto Nazionale di Fisica Nucleare, Sezione di Padova-Trento, $^{bb}$University of Padova, I-35131 Padova, Italy} 

\author{D.~Amidei}
\affiliation{University of Michigan, Ann Arbor, Michigan 48109, USA}
\author{A.~Anastassov}
\affiliation{Northwestern University, Evanston, Illinois 60208, USA}
\author{A.~Annovi}
\affiliation{Laboratori Nazionali di Frascati, Istituto Nazionale di Fisica Nucleare, I-00044 Frascati, Italy}
\author{J.~Antos}
\affiliation{Comenius University, 842 48 Bratislava, Slovakia; Institute of Experimental Physics, 040 01 Kosice, Slovakia}
\author{G.~Apollinari}
\affiliation{Fermi National Accelerator Laboratory, Batavia, Illinois 60510, USA}
\author{J.A.~Appel}
\affiliation{Fermi National Accelerator Laboratory, Batavia, Illinois 60510, USA}
\author{A.~Apresyan}
\affiliation{Purdue University, West Lafayette, Indiana 47907, USA}
% ...
\collaboration{CDF Collaboration\footnote{With visitors from $^a$University of Massachusetts Amherst, Amherst, Massachusetts 01003,
$^b$Istituto Nazionale di Fisica Nucleare, Sezione di Cagliari, 09042 Monserrato (Cagliari), Italy,
$^c$University of California Irvine, Irvine, CA  92697, 
$^d$University of California Santa Barbara, Santa Barbara, CA 93106
$^e$University of California Santa Cruz, Santa Cruz, CA  95064,
$^f$CERN,CH-1211 Geneva, Switzerland,
$^g$Cornell University, Ithaca, NY  14853, 
$^h$University of Cyprus, Nicosia CY-1678, Cyprus, 
$^i$University College Dublin, Dublin 4, Ireland,
$^j$University of Fukui, Fukui City, Fukui Prefecture, Japan 910-0017,
$^k$Universidad Iberoamericana, Mexico D.F., Mexico,
$^l$Iowa State University, Ames, IA  50011,
$^m$University of Iowa, Iowa City, IA  52242,
$^n$Kinki University, Higashi-Osaka City, Japan 577-8502,
$^o$Kansas State University, Manhattan, KS 66506,
$^p$University of Manchester, Manchester M13 9PL, England,
$^q$Queen Mary, University of London, London, E1 4NS, England,
$^r$University of Melbourne, Victoria 3010, Australia,
$^s$Muons, Inc., Batavia, IL 60510,
$^t$Nagasaki Institute of Applied Science, Nagasaki, Japan, 
$^u$National Research Nuclear University, Moscow, Russia,
$^v$University of Notre Dame, Notre Dame, IN 46556,
$^w$Universidad de Oviedo, E-33007 Oviedo, Spain, 
$^x$Texas Tech University, Lubbock, TX  79609, 
$^y$Universidad Tecnica Federico Santa Maria, 110v Valparaiso, Chile,
$^z$Yarmouk University, Irbid 211-63, Jordan,
$^{hh}$On leave from J.~Stefan Institute, Ljubljana, Slovenia, 
}}
\noaffiliation





\begin{abstract}
We report a study of the invariant mass distribution of jet pairs produced in association with a $W$ boson using  data collected with the CDF detector which correspond to an integrated luminosity of 4.3~fb$^{-1}$. The observed distribution has an excess in the 120-160 GeV/c$^2$  mass range which is not described by current theoretical predictions within the statistical and systematic uncertainties. 
In this letter we report studies of the properties of this excess.


\end{abstract}




\maketitle


Measurements of associated production of a $W$ boson and jets are fundamental probes of the electroweak sector of the standard model (SM) and are an essential starting point for searches for physics beyond the SM.
Several important processes share this signature, such as diboson production, associated production of a $W$ and a light Higgs boson and searches for new phenomena \cite{MSSM,techni}.
At the Fermilab Tevatron collider the \DZero collaboration, using a data sample corresponding to an integrated luminosity of 1.1~fb$^{-1}$, reported first evidence for the production of either an additional $W$ or a $Z$ boson in association to a $W$ boson ($WW$ or $WZ$ diboson production) in a lepton plus jets final state~\cite{D0_WWWZ}.
The CDF collaboration recently measured the cross section for the same channel as described in Ref.~\cite{mjjme}.  One of the two methods described in the CDF work uses the invariant mass of the two-jet system ($M_{jj}$) to extract a $WW$ + $WZ$ signal from data.  Here we perform a statistical comparison of that spectrum with expectations by including additional data and further studying the $M_{jj}$ distribution for masses higher than 100 GeV/c$^2$, with minimal changes to the event selection with respect to the previous analysis. We find a statistically significant disagreement with current theoretical predictions. 

The parts of the CDF II detector~\cite{CDFdet}
relevant to this analysis are briefly described here. The tracking
system is composed of silicon microstrip detectors and an open-cell drift
chamber inside a 1.4 T solenoid.  Electromagnetic
lead-scintillator and hadronic iron-scintillator sampling calorimeters
segmented in a projective tower geometry surround the tracking detectors.  A central
calorimeter covers a pseudorapidity range $|\eta| < 1.1$, while ``plug" calorimeters 
extend the acceptance into the region $1.1<|\eta|<3.6$~\cite{coord}.  Outside  the
calorimeters are muon detectors composed of scintillators and drift
chambers.  Cherenkov counters around the beam pipe provide the collider luminosity measurement \cite{lumi}.

The trigger selection used to collect the data sample required a central and high $p_T$ electron (muon).
Further event selection requirements are applied offline to reject backgrounds and
reduce the sensitivity to systematic uncertainties.
We require the presence of one electron (muon) candidate
with $\Et$ ($\Pt$) $>$~20~GeV (GeV/c) and $|\eta|<1.0$ plus missing transverse energy \MET~$>$~25~GeV.
Both electrons and muons are required to be isolated ($Iso<0.1$)~\cite{isolation} to reject leptons from semileptonic decays of heavy 
flavor hadrons and hadrons misidentified as leptons.
Jets are clustered using a fixed-cone algorithm with radius 
$\Delta R = \sqrt{(\Delta\eta)^2+(\Delta\phi)^2} = 0.4$,
and their energies are corrected for detector effects that are of the order of 25\% for jet $\Et = 30$~GeV~\cite{jet_details}~.
Jets with an electron or muon in a cone $\Delta R = 0.52$ around the jet axis are removed. 
Cosmic rays and photon-conversion candidates are  removed.
We require events to have exactly two jets each with $\Et >30$~GeV and $|\eta| < 2.4$, and the 
dijet system to have $\Pt > 40$~GeV/c. 

The transverse mass $M_{T}(W)$~\cite{coord} of the lepton + \MET\ system must be greater
than 30 GeV/c$^2$; the two jets must be separated
by $|\Delta \eta| < 2.5$.
To suppress multijet background, we further require that the direction of \MET~and of the most energetic jet are separated azimuthally by $|\Delta\phi| > 0.4$.

To remove contamination from $Z$ production, we reject events where an additional lepton
 is found using looser criteria and the invariant mass of the two leptons is in the range 76-106 GeV/c$^{2}$.
We further reject events with two identified leptons, where the $\Et$ ($\Pt $) threshold for the second lepton is decreased to 10~GeV~(GeV/c), to suppress other sources of real dileptons such as leptonic decays of both final state $W$'s in $t\bar{t}$  and dibosons with jets. The main difference with respect to the selection criteria used in Ref.~\cite{mjjme} is that the jet $E_T$ threshold is increased from 20~GeV to 30~GeV, motivated by the interest in a higher invariant mass range. This analysis critically depends on the shape of the steeply falling dijet mass distribution. For this reason, we verified by Monte Carlo studies that our selection does not sculpt the dijet invariant mass distribution of any process expected to contribute to the sample at masses above 100 GeV/$c^2$.
The resulting sample is dominated by events where a $W$ boson, which decays leptonically, is produced 
in association with jets ($W$+jets).  Minor contributions to the selected sample
come from $WW$+$WZ$, $t\bar{t}$, $Z$+jets, single top production and multijet QCD sources. 
Predictions for these processes, with the exception of the multijet QCD component,  are obtained using event
generators and a {\sc Geant}-based CDF II detector simulation~\cite{cdfsim}.  The diboson, $t\bar{t}$, and single top
components are simulated using the {\sc Pythia} event generator~\cite{pythia}.  
The $W$+jets and $Z$+jets processes are simulated using a matrix element Leading Order event generator  {\sc Alpgen}~\cite{alpgen} with an interface to 
{\sc Pythia} providing parton showering and hadronization~\cite{MLM,tesi}.
Multijet QCD events, where one jet is misidentified as a lepton, are modeled with data containing
anti-isolated muons ($Iso>0.2$) or candidate electrons failing quality cuts~\cite{tesi}.
The normalization of the $Z$+jets component is based on the measured cross 
section~\cite{mesxs}, while for $t\bar{t}$, single top, and diboson production the NLO predicted 
cross sections are used~\cite{theoxs}.  The detection efficiencies for $Z$+jets, $t\bar{t}$, 
single top, and diboson contributions are determined from simulation.
The normalization of the multijet QCD component and a preliminary estimation of the $W$+jets component are
obtained by fitting the $\MET$ spectrum in data to the sum of all
contributing processes.

We perform a combined binned $\chi^{2}$ fit, for electron and muon events, to the dijet invariant mass ($M_{jj}$) spectrum 
using predictions for the multijet QCD, $WW$, $WZ$, $Z$+jets, $W$+jets, $t\bar{t}$, and single top
processes.  
The final $W$+jets normalization is determined by minimizing this $\chi^{2}$ and all other contributions are constrained to be within the variance of their expected normalization.

\begin{figure*}[ht]
\includegraphics[angle=90,width=0.48\textwidth]{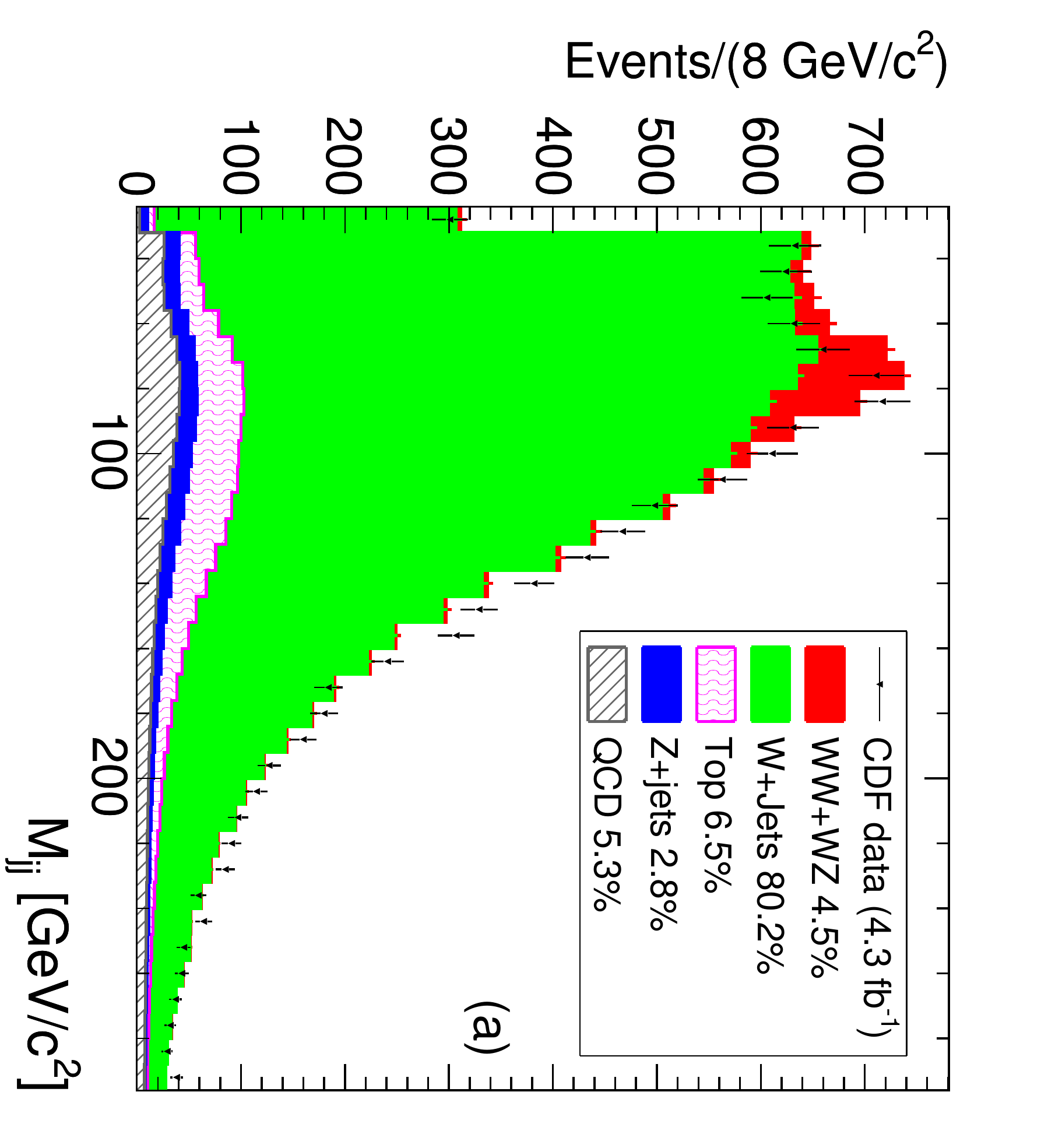}
\includegraphics[angle=90,width=0.48\textwidth]{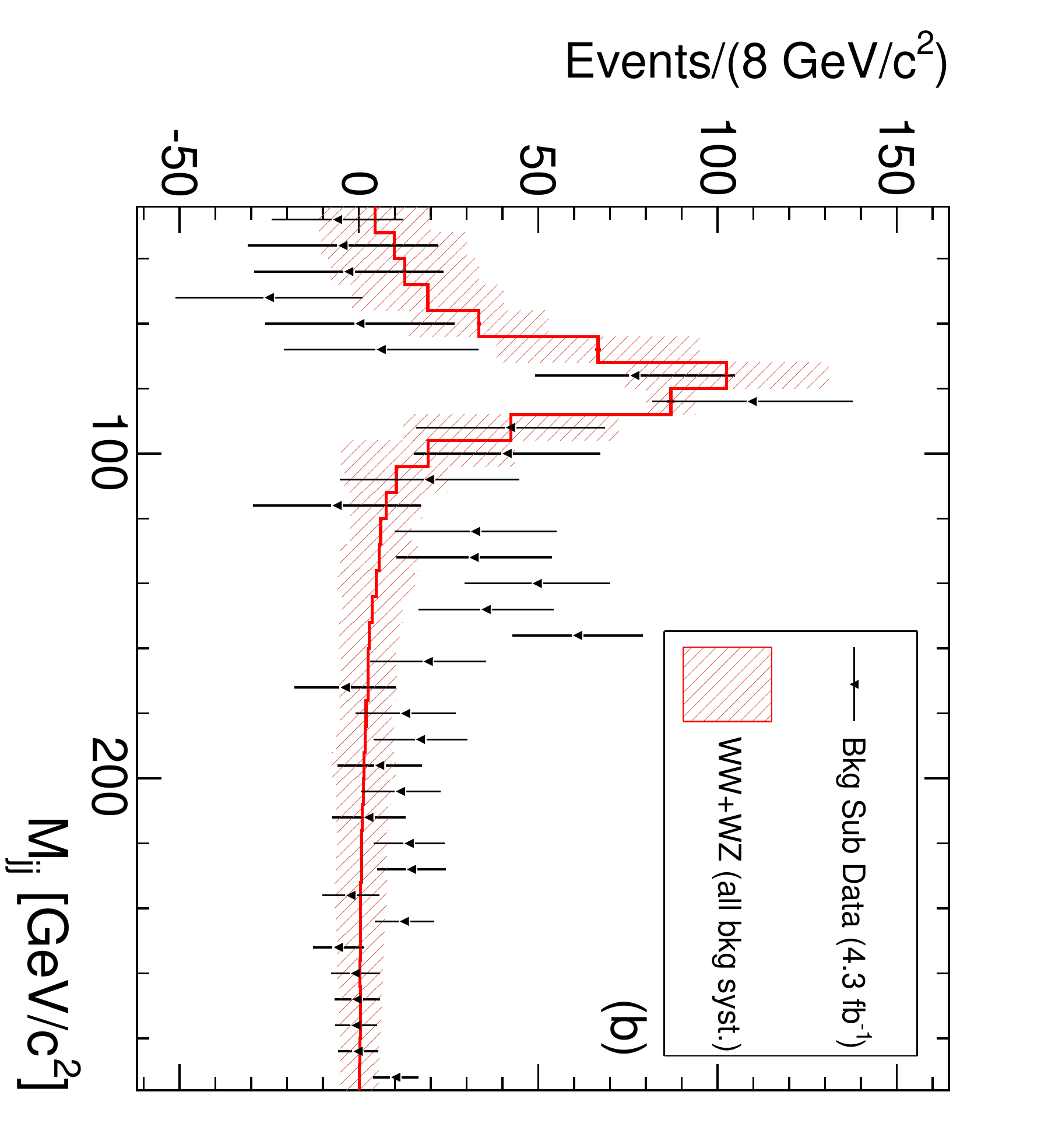} \\
\includegraphics[angle=90,width=0.48\textwidth]{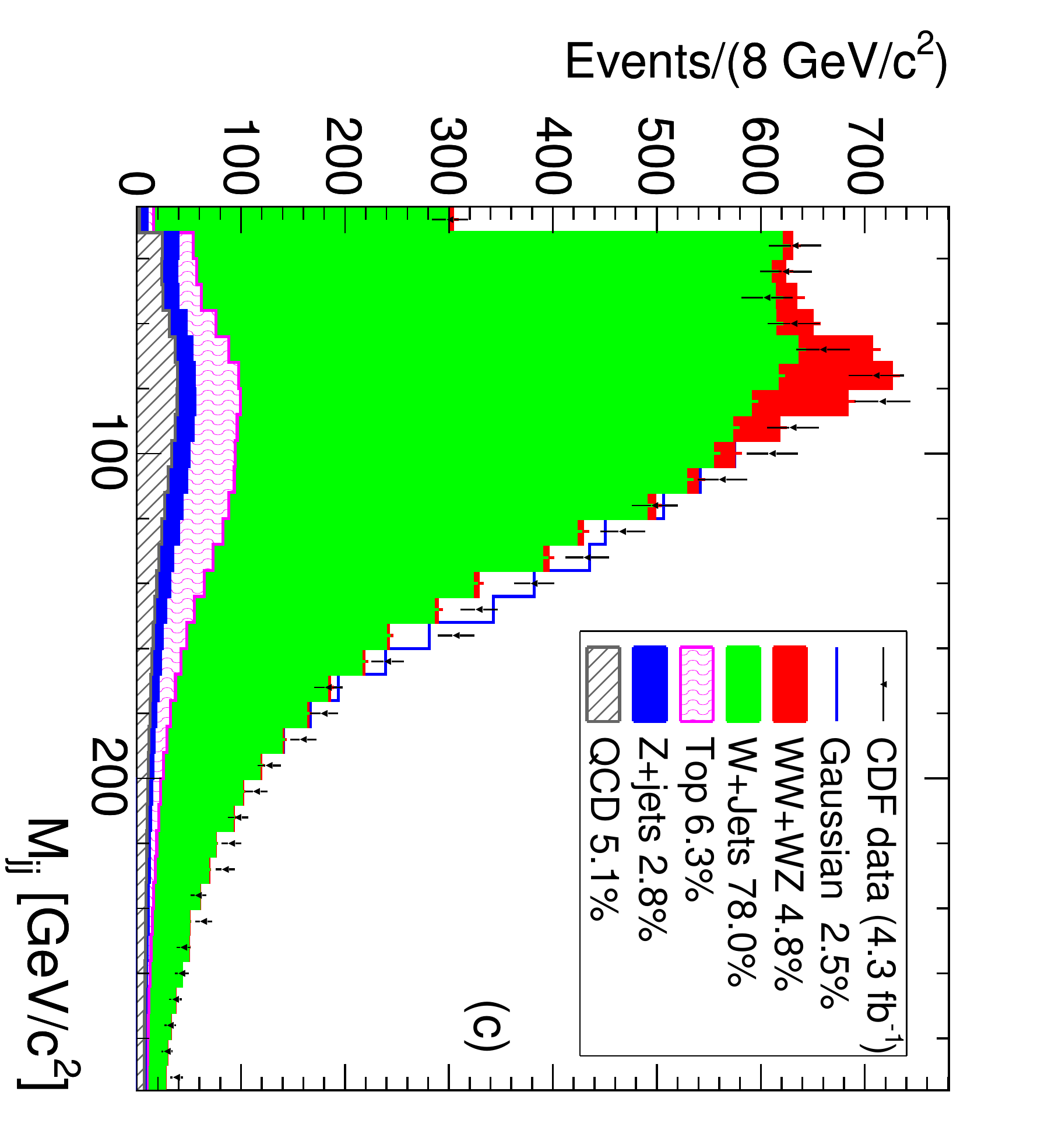}
\includegraphics[angle=90,width=0.48\textwidth]{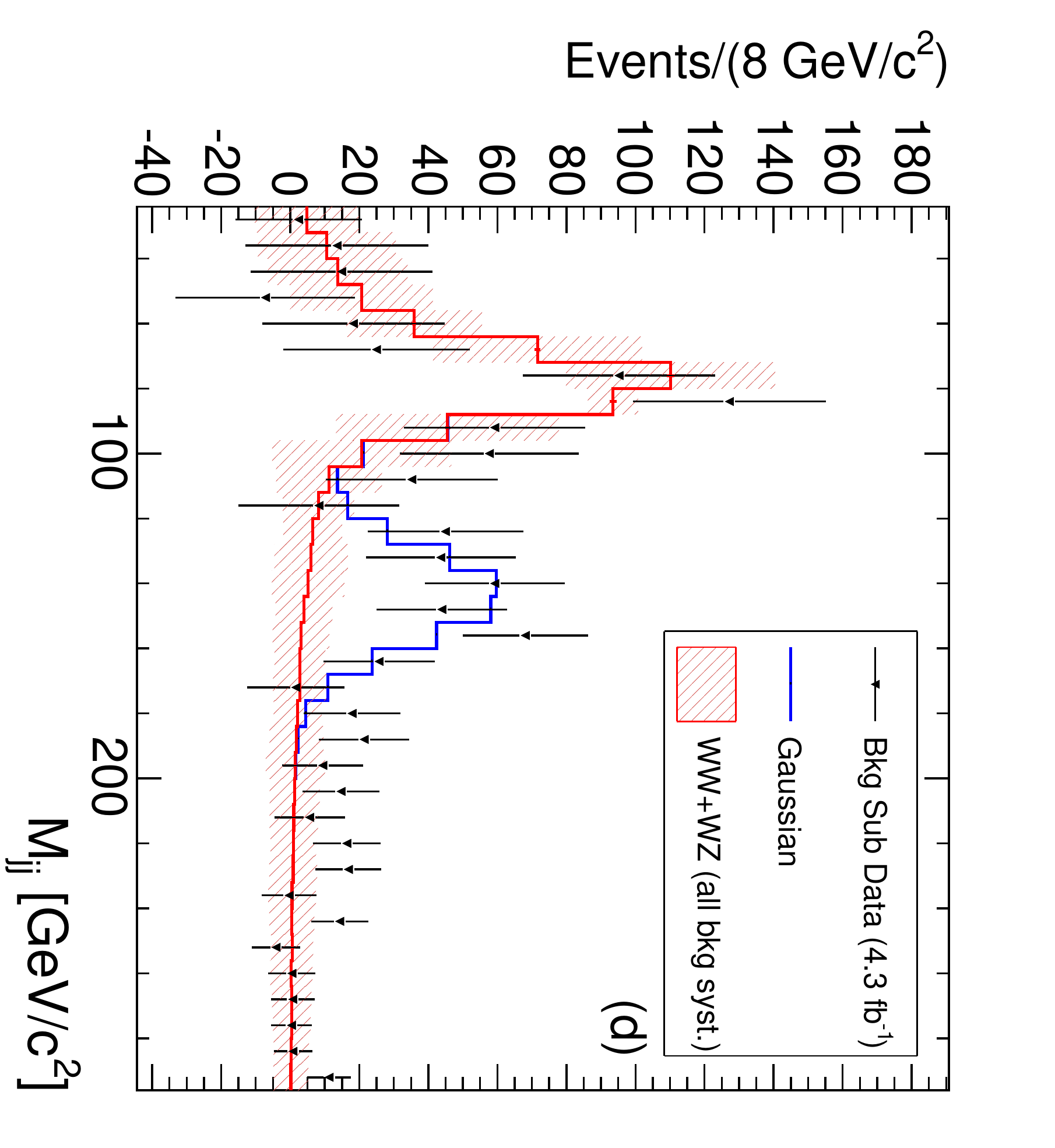}
\caption{\label{fig:projections} 
The dijet invariant mass distribution.
The sum of electron and muon events is plotted.
In the left plots we show the fits  for known processes only (a)
and with the addition of a hypothetical Gaussian component (c).
On the right plots we show, by subtraction, only the resonant contribution to $M_{jj}$ including
$WW$ and $WZ$ production (b) and the hypothesized narrow Gaussian contribution (d).
In plot (b) and (d) data points differ because the normalization of the background changes between the two fits.
The band in the subtracted plots represents the sum of all background shape systematic uncertainties described in the text.
The distributions are shown with a 8 GeV/c$^{2}$ binning while the actual fit is performed using a 4 GeV/c$^{2}$ bin size.
}
\end{figure*}

We fit the dijet mass distribution in the range 28-200~GeV/c$^2$ defined \textit{a priori} in
the measurement of the $WW/WZ$ cross section \cite{mjjme}.  Figs.~\ref{fig:projections} (a) and (b) show the extrapolation
of this fit in the extended range of mass up to 300~GeV/c$^2$.
The fit is stable with respect to changes in the fit range and histogram binning.
Our model describes the data within uncertainties, except in the mass region $\sim$ 120-160~GeV/c$^{2}$, where an excess over the simulation is seen.  
The fit $\chi^2$/ndf is 77.1/84, where ndf is the number of degrees of freedom.  The $\chi^2$/ndf computed only in the region 120-160~GeV/c$^2$ is 26.1/20.
However the Kolmogorov-Smirnov (KS) test, which is more sensitive to a localized excess, yields a probability of $6 \times 10^{-5}$~\cite{KStest}.

We try to model the excess with an additional Gaussian peak and perform a $\Delta \chi^{2}$ test of this hypothesis.
The Gaussian is chosen as the simplest hypothesis compatible with the assumption of a two jet decay of a narrow resonance with definite mass.
The width of the Gaussian is fixed to the expected dijet mass resolution by scaling the width of the $W$ peak in the same spectrum: $\sigma_{\rm{resolution}} = \sigma_{W} \sqrt{\frac{M_{jj}}{M_{W}}}= 14.3$ GeV/c$^2$, where $\sigma_{W}$ and ${M_{W}}$  are the resolution and the average dijet invariant mass for the hadronic $W$ in the $WW$ simulations respectively,  and ${M_{jj}}$ is the dijet mass where the Gaussian template is centered.

In the combined fit, the normalization of the Gaussian is free to vary independently 
for the electron and muon samples, while the mean is constrained to be the same.
The result of this alternative fit is shown in  Figs.~\ref{fig:projections} (c) and (d).
The inclusion of this additional component brings the fit into good agreement with the data.
The fit $\chi^2$/ndf is 56.7/81 and the Kolmogorov-Smirnov test returns a probability of 0.05, accounting only for statistical uncertainties.
The $W$+jets normalization returned by the fit including the additional Gaussian component is compatible with the preliminary estimation from the $\MET$ fit. 
The $\chi^2$/ndf  in the region 120-160~GeV/c$^2$ is 10.9/20.
The values of parameters returned by the combined fit are shown in Table~\ref{tab:yields}, 
where the mean of the Gaussian peak represents the experimentally measured value i.e. it is not corrected back to the parton-level. 

\begin{table}[t]
\caption{\label{tab:yields}
Results of the combined fit. The ratios of the number of
events in the excess to the number of expected diboson
events in the electron and muon samples are statistically compatible with each other.}

\begin{ruledtabular}
\begin{tabular}{lcc}
                                 & Electrons & Muons \\ \hline
Excess events             & $156\pm42$ & $97\pm38$ \\ 
Excess events / expected diboson & $0.60 \pm 0.18$ & $0.44 \pm 0.18$ \\ 
Mean of the Gaussian component                  & \multicolumn{2}{c}{$144 \pm 5$~GeV/c$^2$} \\
\end{tabular}
\end{ruledtabular}
\end{table}

We take the difference between the  $\chi^{2}$ of the two fits ($\Delta \chi^{2}$), 
with and without the additional Gaussian structure 
to assess the significance of the excess.
The expected distribution of $\Delta \chi^{2}$ is computed numerically from simulated background-only experiments and used to derive the p-value corresponding to the  $\Delta \chi^{2}$ actually observed.
In order to account for the trial factor within our search window, 120-200~GeV/c$^2$, in each pseudoexperiment we calculate the $\Delta \chi^{2}$ varying the position of the Gaussian component in steps of 4~GeV/c$^2$. The largest $\Delta \chi^{2}$ for each pseudoexperiment is used to define the p-value distribution.

In deriving the p-value we account for systematic uncertainties that affect the background shapes and the normalization of constrained components.
Normalization uncertainties of unconstrained components are considered as part of the statistical uncertainty.
The largest systematic uncertainties arise from the modeling of the $W$+jets and multijet QCD  shapes.
For $W$+jets we consider, as an alternative, the $M_{jj}$ distributions obtained by halving or doubling the renormalization scale ($Q^2$) in {\sc Alpgen}.
For multijet QCD, we change our model using different lepton isolation ranges.
The systematic uncertainty due to uncertainties in the jet energy scale ($\pm 3$\%) affects all components with the exception of multijet QCD, which is derived from data.
For each systematic effect we consider the two extreme cases. For each of the possible combinations of systematic effects we calculate a different $\Delta \chi^{2}$ distribution and take the conservative approach of using the distribution that returns the highest p-value. The total systematic effect on the extracted number of excess events, defined as the number of events fitted by the Gaussian component, in the electron and muon samples is found to be 10\% and 9\%, respectively. The dominant systematic effects arise from the $W$+jets renormalization scale (6.7\%), the jet energy scale (6.1\%) and QCD shape (1.9\%).
Assuming only background contributions, and systematic errors, the probability to observe an excess larger than in the data is $7.6 \times 10^{-4}$ corresponding to a significance of 3.2 standard deviations for a Gaussian distribution. 
For comparison, the p-value without taking into account systematic uncertainties is $9.9 \times 10^{-5}$.
\begin{figure*}[ht]
\includegraphics[angle=0,width=0.48\textwidth,height=0.4\textwidth]{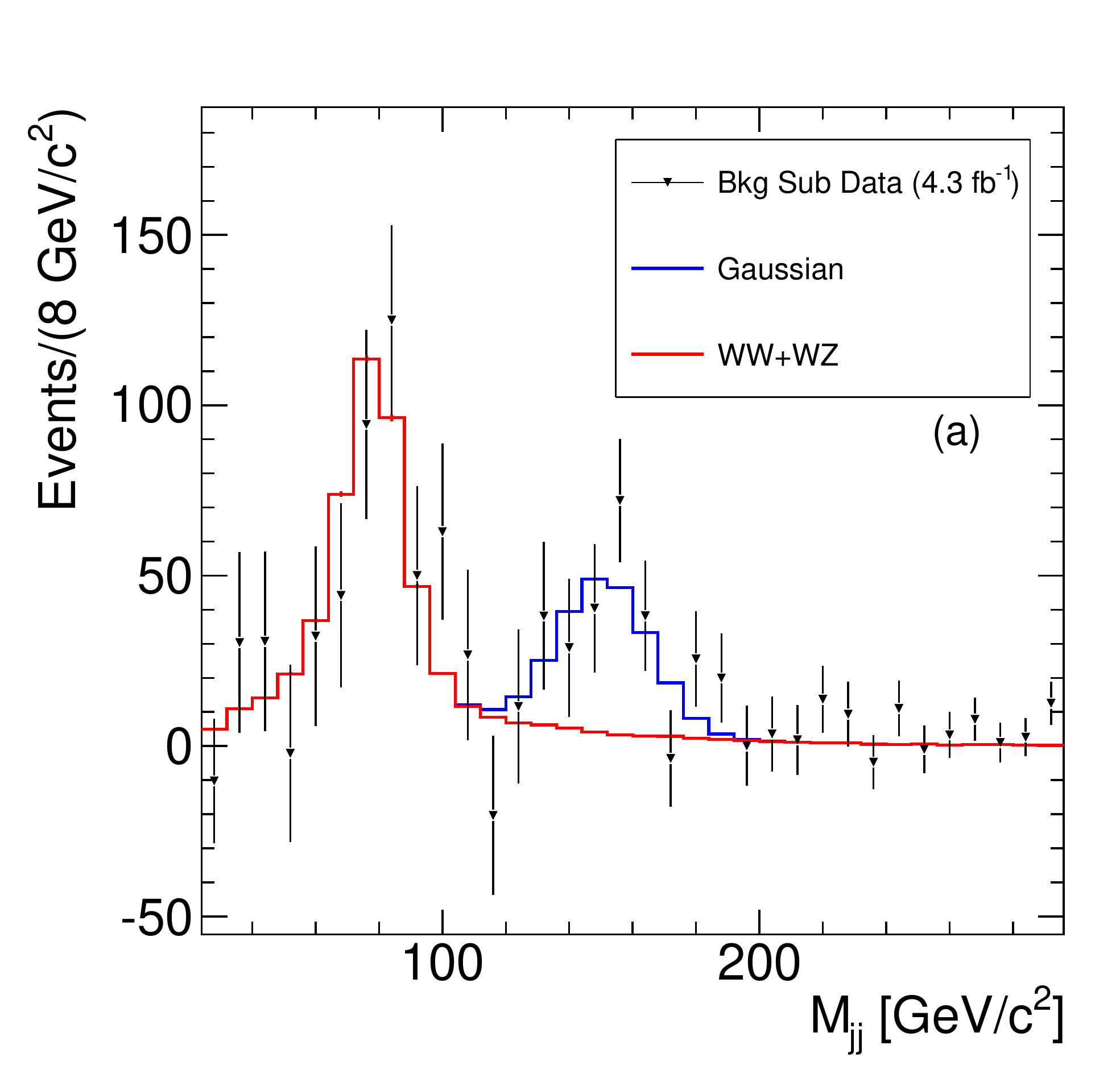}
\includegraphics[angle=0,width=0.48\textwidth,height=0.4\textwidth]{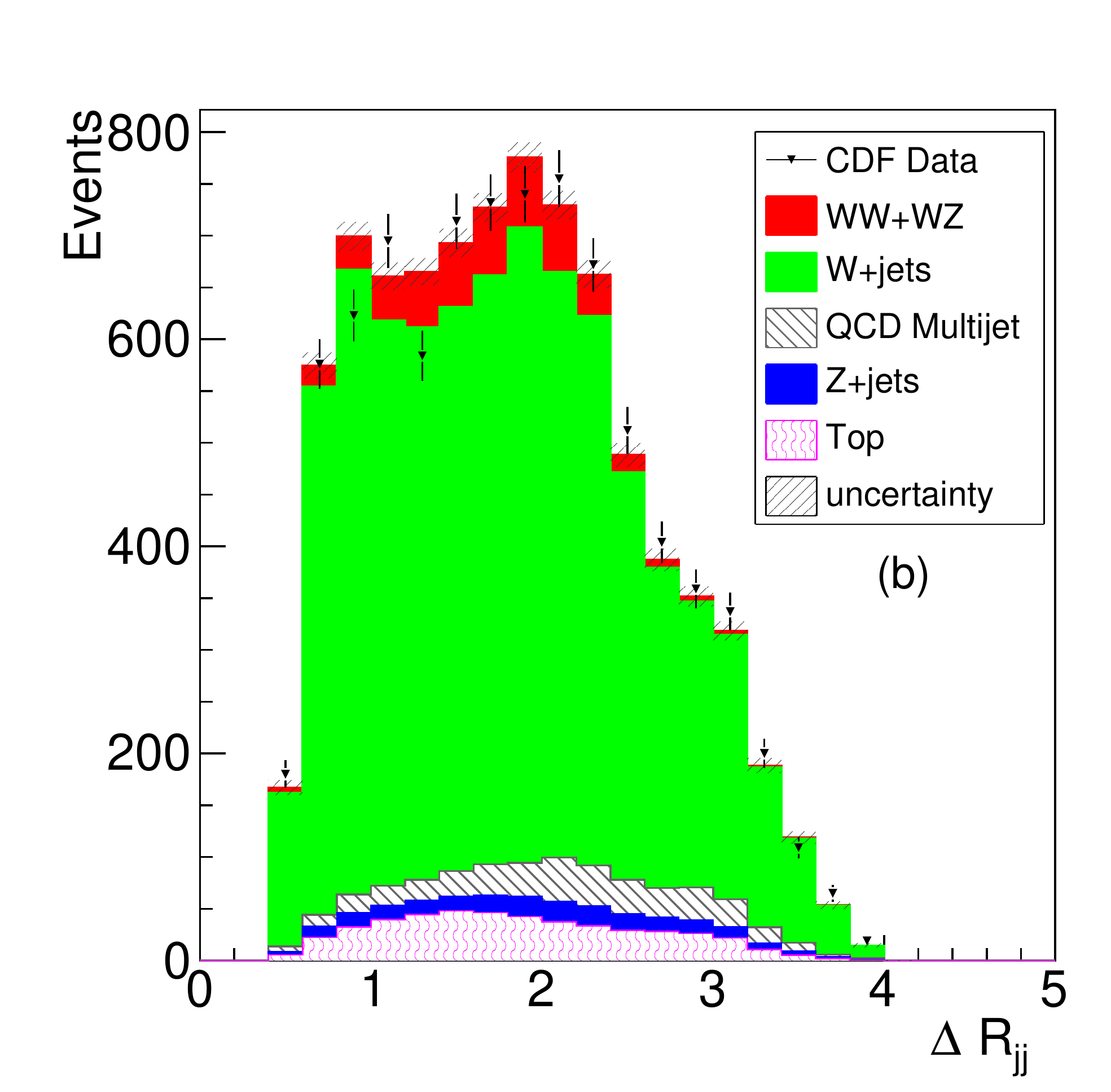}

\caption{\label{fig:Q2} 
The dijet invariant mass distribution for the sum of electron and muon events is shown after subtraction of fitted background components
 with the exception of resonant contribution to $M_{jj}$ including $WW$ and $WZ$ production and the hypothesized narrow Gaussian contribution (a).
With respect to Figure~1, the subtracted background components are chosen as the systematic combination that best fit data (see text). The fit $\chi^2$/ndf is 62.0/81. (b) $\Delta R_{jj}$ distribution for events with $M_{jj}<$ 115 and $M_{jj}>$ 175 GeV/c$^{2}$ of the data compared to the background estimation that corresponds to the same systematic combination of~(a). The uncertainty band corresponds to background statistical uncertainty.
 }
\end{figure*}

To investigate possible mismodeling of the $W$+jets background we consider various configurations of our systematic uncertainties.
The combination of systematic uncertainties that fits the data best is shown in Fig.~\ref{fig:Q2} (a) where Q$^2$ is doubled  and the QCD shape is varied. The KS probability for this fit is 0.28. The fit $\chi^2$/ndf outside the 120-160 GeV/c$^2$ region is 50.3/66, indicating that the dijet mass distribution is well modeled within our systematic uncertainties.
This choice of systematic uncertainties returns a p-value intermediate between the central configuration and the most conservative combination.
In order to test ``Next to Leading Order'' contributions to the $W$+2 partons prediction, 
we compare a sample of $W$+2 partons simulated with {\sc Alpgen} and interfaced to {\sc Pythia} for showering to a sample of $W$+2 partons simulated using the MCFM generator~\cite{mcfm1}.
We extract a correction as a function of $M_{jj}$ that is applied to the {\sc Alpgen} + {\sc Pythia} sample used in our background model.
The statistical significance obtained with the MCFM reweighted $W$+jets background model is 3.4$\sigma$. 

Details of a large set of additional checks can be found in Ref.~\cite{tesi}. In particular we verified that the background model describes the data in several independent control regions and satisfactorily reproduces the kinematic distributions  of jets, lepton, and $\met$. 
The excess is stable against 5 GeV variations of the thresholds used for all of the kinematic selection variables, including  variations of the jet $E_T>30$~GeV threshold. 
This analysis employs requirements on jets of $E_T >$ 30 GeV and  $\Pt > 40$~GeV/c for the dijet system, which improves the overall modeling of many
kinematic distributions.  We also test a selection only requiring jet $E_T > 20$ GeV as in Ref.~\cite{Higgs}. This selection, which increases the background by
a factor of 4, reduces the statistical significance of the excess to about 1$\sigma$.

We study the $\Delta R_{jj}$ distribution to investigate possible effects that could result in a mismodeling of the dijet invariant mass distribution.
We consider two control regions, the first defined by events with $M_{jj}<$ 115 and $M_{jj}>$ 175 GeV/c$^{2}$ and the second defined by events with $\Pt < 40$~GeV/c. 
We use these regions to derive a correction as a function of $\Delta R_{jj}$ to reweight the events in the excess region. We find that the reweightings change the statistical significance of the result by plus or minus one sigma. 
However, the $\Delta R_{jj}$ distribution is strongly correlated to $M_{jj}$ and the control regions both have significantly different distributions of $\Delta R_{jj}$. Reweighting our $W$+jets sample to correct for the differences observed in  $\Delta R_{jj}$ in the control samples may be indicative of the effect of correcting $\Delta R_{jj}$  mismodeling or may introduce bias in the $M_{jj}$ distribution.  In addition, the $\Delta R_{jj}$ distribution is consistent within the one sigma variation of the systematic uncertainties for events outside the excess mass region as shown in Fig.~\ref{fig:Q2} (b). The data-background comparison of the $\Delta R_{jj}$ distribution has $\chi^2$/ndf of 26.7/18 and a KS probability of 0.022 when compared with best-fit systematic model.
For these reasons, we present these studies as cross checks and quote the significance in the unweighted sample as our primary result.

We look for evidence in favor or against the hypothesis that the excess in the 120-160~GeV/c$^2$ mass range is from a new (non-SM) physics source.
Since non-SM particles may in general couple to both massive electroweak gauge bosons we have investigated the shape of the dijet mass
distribution in $Z$+jets events.  
In this sample the number of events in the data is approximately a factor 15 less than in the $W$+jets sample and no statistically significant deviation from the SM expectation is observed.
We increase the jet $\Et$ threshold in steps of 5~GeV and check the fraction of excess events that are selected as a function of the jet $\Et$. 
The result is compatible with expectation from a Monte Carlo simulation of a $W$ boson plus a  particle with a mass of 150~GeV/c$^{2}$ and decaying into two jets~\cite{tesi}. 
In this model, we estimate a cross section times the particle branching ratio into dijets of the order of 4~pb.
The cross section of the observed excess is not compatible with SM $WH$ production whose  $\sigma \cdot BR(H\rightarrow b\bar{b})$ is about 12~fb for $m_H=150$~GeV/c$^{2}$~\cite{Higgs_xsec_and_BR}.  
To check the flavor content with this selection, 
we identify jets originating from a b-quark by requesting a displaced secondary vertex for tracks within the jet cone. We compare the fraction of events with at least one b-jet in the excess region  (120-160~GeV/c$^{2}$) to that in the sideband regions (100-120 and 160-180~GeV/c$^{2}$), and find them to be compatible with each other.
Dedicated CDF searches for $WH \rightarrow l\nu b\bar{b}$ using events with reconstructed  displaced vertices from b hadron decay, and looser selection criteria,  have not found any significant excesses using final analysis discriminants trained to identify Higgs bosons in the mass range 100-150 GeV/c$^2$~\cite{Higgs}. 

Finally, to investigate the possibilities of a parent resonance or other quasi-resonant behavior, we consider 
the $M_{({\rm lepton},\nu,jj)}$ and the $M_{({\rm lepton},\nu,jj)}-M_{jj}$~\cite{M_lnujj} distributions for events with $M_{jj}$ in the range 120-160~GeV/c$^{2}$
and,  to investigate the Dalitz structure of the excess events, the distribution of $M_{({\rm lepton},\nu,jj)}-M_{jj}$, in bins of $M_{jj}$.
The distributions are compatible in shape with the background-only hypothesis in all cases. 

In conclusion, we study the invariant mass distribution of jet pairs produced in association with a $W$ boson. 
The best fit to the observed dijet mass distribution using known components, and modeling the dominant $W$+jets background using {\sc Alpgen}+{\sc Pythia} Monte Carlo, shows a statistically significant disagreement. One possible way to interpret this disagreement is as an excess in the 120-160~GeV/c$^{2}$ mass range. If we model the excess as a Gaussian component with a width compatible with the dijet invariant mass resolution, and perform a $\Delta \chi^{2}$ test for the presence of this additional component, we obtain a p-value of $7.6 \times 10^{-4}$, corresponding to a significance of 3.2 standard deviations, after accounting for all statistical and systematic uncertainties.

We thank the Fermilab theory group for helpful suggestions, particularly J.~M.~Campbell, E.~J.~Eichten, R.~K.~Ellis, C.~T.~Hill, and A.~O.~Martin. We are grateful to K.~D.~Lane and M.~L.~Mangano. We thank the Fermilab staff and the technical staffs of the participating institutions for their vital contributions. This work was supported by the U.S. Department of Energy and National Science Foundation; the Italian Istituto Nazionale di Fisica Nucleare; the Ministry of Education, Culture, Sports, Science and Technology of Japan; the Natural Sciences and Engineering Research Council of Canada; the National Science Council of the Republic of China; the Swiss National Science Foundation; the A.P. Sloan Foundation; the Bundesministerium f\"ur Bildung und Forschung, Germany; the Korean World Class University Program, the National Research Foundation of Korea; the Science and Technology Facilities Council and the Royal Society, UK; the Institut National de Physique Nucleaire et Physique des Particules/CNRS; the Russian Foundation for Basic Research; the Ministerio de Ciencia e Innovaci\'{o}n, and Programa Consolider-Ingenio 2010, Spain; the Slovak R\&D Agency; and the Academy of Finland.


\begin{thebibliography}{99}
\bibitem{coord} We use a cylindrical coordinate system with its origin in the center 
of the detector, where $\theta$ and $\phi$ are the polar and azimuthal angles, respectively,
and pseudo-rapidity is $\eta= - \rm ln$ $\rm tan(\theta/2)$. 
The transverse energy $E_T$ (momentum $p_T$) is defined as $E\sin\theta$ ($p\sin\theta$).
The missing $E_T$ ($\METVEC$) is defined by $\METVEC = -\sum_iE_T^i{\hat n}_i$,
where ${\hat n}_i$ is a unit vector perpendicular to the beam axis and pointing at the $i^{th}$ calorimeter tower.  
~$\METVEC$ is corrected for high-energy muons and also jet energy corrections.
We define $\MET=|\METVEC|$. The transverse mass of the $W$ is defined as $M_T (W)=\sqrt{2p_T^l \MET (1-\cos(\Delta \phi^{l\nu}))}$.
\bibitem{isolation} Lepton isolation ($Iso$) is defined as $\frac{\sum E_T}{E_T}$, $\frac{\sum E_T}{p_T}$ for electrons and muons respectively, where $\sum E_T$ is the calorimetric energy in a cone 0.4 around the lepton.
\bibitem{KStest} The reported KS probability corresponds to the KS test between data and background distributions. It does not account for the fact that the background distributions are constrained by fits to data. The reported values are thus an upper limit on KS probability.
\bibitem{M_lnujj} $M_{({\rm lepton},\nu,jj)}$ denotes the total invariant mass of the lepton, neutrino and dijet system. We reconstruct the longitudinal component of the neutrino momentum by imposing a $W$ mass of 80.398~GeV/$c^2$. We consider both real solutions for the $p_z$ of the neutrino  and we discard complex solutions of the $W$ mass equation.
\end{thebibliography}
\end{document}